# Gate tunable spin-orbit coupling and weak antilocalization effect in an epitaxial La$_{2/3}$Sr$_{1/3}$MnO$_3$ thin film

Shao-Pin Chiu[1,*], Michihiko Yamanouchi[2,3], Tatsuro Oyamada[3], Hiromichi Ohta[2,3], and Juhn-Jong Lin[1,4]

[1]*Institute of Physics, National Chiao Tung University, Hsinchu 30010, Taiwan*

[2]*Research Institute for Electronic Science, Hokkaido University, N20W10, Kita, Sapporo 001−0020, Japan*

[3]*Institute of Materials Research, University of Tokyo, Bunkyo, Tokyo 123-4567, Japan*

[4]*Department of Electrophysics, National Chiao Tung University, Hsinchu 30010, Taiwan*

[*]E-mail: fluentbb@gmail.com

**Abstract:**

Epitaxial La$_{2/3}$Sr$_{1/3}$MnO$_3$ (LSMO) films have been grown on SrTiO$_3$ (001) substrates via pulsed laser deposition. In a 22-nm thick LSMO film with a low residual resistivity of $\rho_0 \simeq 59$ μΩ cm, we found a zero-field dip in the magnetoresistance (MR) below 10 K, manifesting the weak antilocalization (WAL) effect due to strong spin-orbit coupling (SOC). We have analyzed the MR data by including the D'yakonov-Perel' spin-relaxation mechanism in the WAL theory. We explain that the delocalized spin-down electron subband states play a crucial role for facilitating marked SOC in clean LSMO. Moreover, we find that the SOC strength and gate voltage tunability is similar to that in the 2DEG at LaAlO$_3$/SrTiO$_3$ interface, indicating the presence of an internal electric field near the LSMO/SrTiO$_3$ interface. In a control measurement on a 5-nm thick high resistivity ($\rho_0 \simeq 280$ μΩ cm) LSMO film, we observe only a small zero-field peak in MR from weak localization effect, indicating negligible SOC.

**I. INTRODUCTION**

The perovskite manganite La$_{2/3}$Sr$_{1/3}$MnO$_3$ (LSMO) has recently been exploited for its possible use in spintronics. Multiple researchers have found a nearly 100% spin polarization (*P*) in LSMO films from the experiments of spin-resolved photoemission spectroscopy [1] and magnetoresistance (MR) in magnetic tunnel junctions [2]. They have ascribed the results to intrinsic half-metallicity and classified LSMO as a traditional, type I half metal. On the other hand, point-contact Andreev reflection (PCAR) studies showed a broad range of *P* (58–92%) in electrical current [3]. Moreover, the *P* and residual resistivity revealed a correlation demonstrating that highly resistive samples exhibited higher *P* values [3]. At first glance, this is a surprising correlation opposite to what people would expect for a traditional half metal. This puzzle can be understood in terms of the energy bandstructure of a type III$_A$ half metal, as schematically depicted in Fig. 1(a) [4]. The energy bandstructure of Fig. 1(a) constitutes of a spin-up subband of relatively mobile holes and a spin-down subband of comparatively heavy electrons. There are theoretical calculations [5,6] and experiments [7,8] on LSMO, which are supportive of this kind of bandstructure. In this type of half metals, there is no energy gap between the two 3*d* spin subbands. In the presence of strong disorder, the spin-down subband states can become (largely) localized, and hence the electrical transport properties are governed entirely by the spin-up subband states [6]. In a type I half metal, scattering between spin-up carriers and spin-down carriers will mostly be frozen out at low temperatures, due to an energy gap between the two subbands. Therefore, one does not expect (strong) spin-relaxation process to occur even if the spin-orbit coupling (SOC) strength is finite. On the contrary, in a type III$_A$ half metal, spin-relaxation processes can become marked if the spin-down subband states are delocalized. Then, interaction between spin-up holes and spin-down electrons can take place, which may foster fast spin relaxation of the charge carriers. Microscopically, the spin-relaxation interaction can be substantiated by a finite SOC which prevails in the material/device under study. A clean LSMO thick film with delocalized spin-down subband states provides an opportunity to test this concept. The SOC originates from an internal electrical field which is induced at the LSMO/ SrTiO$_3$ (LSMO/STO) interface.

In low-dimensional systems, the low-field MR due to the weak localization (WL) and weak antilocalization (WAL) effect provides a powerful tool for extracting the spin-orbit scattering time, $\tau_{so}$, and the corresponding SOC splitting $\Delta_{so}$ (defined below) [9,10]. In the presence of SOC, the spin-part wavefunction will change sign over a characteristic length scale called the spin-orbit scattering length $L_{so} = \sqrt{D\tau_{so}}$, where $D$ is the charge carrier diffusion constant. When $L_{so}$ is much shorter than the electron (hole) dephasing length $L_{\varphi} = \sqrt{D\tau_{\varphi}}$, where $\tau_{\varphi}$ is the electron (hole) dephasing time, the WAL effect with a zero-field dip in MR is expected. In the opposite limit of negligible SOC ($L_{so} \gg L_{\varphi}$), the WL effect with a zero-field peak in MR is expected. Thus far, only the WL effect has been observed in a 10-nm thick epitaxial LSMO film by Niu *et al*. [11]. In this paper, we report our experimental realization of the WAL effect in a 22-nm thick epitaxial LSMO film. Our film is relatively clean and has a residual resistivity 30 times lower than that of the 10-nm sample reported in Ref. [11]. In particular, the backgate-voltage, $V_{bg}$, tunability of $\Delta_{so}$ found in this film indicates that the SOC originates from an interfacial Rashba-type interaction. Therefore, the two spin subbands nearby the LSMO/STO interface split along the momentum direction owing to the Rashba effect [12], apart from the magnetism induced double exchange splitting along the energy direction [Fig. 1(a)]. Away from the interface, only the exchange splitting remains effective in the bulk of LSMO film.

## II. EXPERIMENTAL METHOD

LSMO films with nominal thickness of 5 and 22 nm were heteroepitaxially grown on (001) SrTiO$_3$ single-crystalline substrates by pulsed laser deposition (PLD) at the substrate temperature of 700°C under oxygen atmosphere ($P_{O2}$ = 25 Pa). A KrF excimer laser ($\lambda$ = 248 nm, pulse duration ~20 ns, fluence ~1.6 J cm$^{-2}$ pulse$^{-1}$, 1 Hz, COMPex 102) was used to ablate the ceramic target of LSMO. After the deposition, the LSMO films were annealed for 20 min under the identical conditions for the growth. The thickness of the resultant films was measured using X-ray reflectivity (XRR, ATX-G, Rigaku Co.) with monochromated Cu K$\alpha_1$ radiation. The resistance and MR were measured with ac resistance bridges (Linear Research model LR700 or LR400 operating at 16 Hz), by employing the van der Pauw electrode configuration. The backgate voltage was applied by a Keithley model 2635A sourcemeter. Low temperature measurements were performed on an Oxford Heliox $^3$He cryostat equipped with a 2-Tesla superconducting magnet. In the following, we focus mostly on the 22-nm thick epitaxial LSMO film.

## III. RESULTS AND DISCUSSION

Figure 1(b) shows the temperature dependence of resistivity, $\rho(T)$, for the 22-nm thick LSMO film between 0.36 and 300 K. [The $\rho(T)$ for the 5-nm thick film is also plotted for comparison.] A large relative resistance ratio of $\rho(290\ \text{K})/\rho(10\ \text{K}) = 17$ indicates the sample being a good metal. The residual resistivity is $\rho_0 = 58.7$ μΩ cm. This $\rho_0$ value is as low as that of the optimal samples fabricated by PLD method [13] and close to that of the cleanest samples used in PCAR experiments [3]. Therefore, the lattice structure of this film is of high quality, with a low defect number density and minimal grain boundaries.

It has previously been found that, in LSMO films grown on various substrates, an interfacial, insulating ("dead") layer often exists. The thickness of the insulating layer varies from 2.4 to 5 nm [14,15]. If taking this matter into account, the effective thickness (and the $\rho_0$ value given above) of our film will be reduced by 10–20%.

Figure 2(a) shows the sheet resistance $R_s$ as a function of magnetic field $H$ at several temperatures, as indicated. The magnetic field was applied perpendicular to the film plane. We first examine the overall MR behavior in the wide magnetic field range of $|H_c| \leq 1.2$ T. Previously, a large negative MR due to grain boundary scattering has been found in polycrystalline LSMO films [16,17]. In contrast, here we observe a small positive MR which is similar to that seen in thick, clean epitaxial LSMO in [17]. In the field range $|H| < 0.7$ T, hysteretic behavior is observed, which stems from the alignment processes of magnetic domains in ferromagnetic LSMO. In Fig. 2(b), $H_{c+}$ and $H_{c-}$ denote the coercive fields, with the value $|H_c| \approx 0.27$ T. In the rest of this paper, we shall focus on the low magnetic field regime of $H \ll |H_c|$ to address the WAL MR. In Fig. 2(b), the $R_s(H)$ curve reveals an evident zero-field dip, or a zero-field peak in the sheet magneto-conductance $G_s(H) = 1/R_s(H)$, as shown in the inset. The magnitude of the zero-field peak in $G_s(H)$ amounts to about one half of the quantum conductance $e^2/h$, where $e$ is the electronic charge, and $h$ is the Planck constant. Figure 2(a) demonstrates that the zero-field dip decreases with increasing temperature $T$, and it vanishes at ≈10 K [see also Fig. 4(a)]. These temperature dependent positive MR are the quantum-interference manifestations of the WAL effect induced by marked SOC [18].

We evaluate the charge carrier elastic mean free time to be $\tau_e \approx 2.9$ fs, using the Drude model with our measured $\rho_0$ value, a carrier density $n$ extracted from the Hall effect measurement ($1.25 \times 10^{22}$ cm$^{-3}$ in a 25-nm thick LSMO on STO) [19], and an effective mass of the majority carrier (hole) $m^* = 0.6 m_0$ [6,19], where $m_0$ is the free electron mass. From the free-electron model, we estimate the carrier elastic mean free path to be $l_e \simeq 4$ nm, the diffusion coefficient $D = v_F l_e / 3 \simeq 18.5$ cm$^2$/s, and the product $k_F l_e \simeq 28$, where $v_F$ is the Fermi velocity, and $k_F$ is the Fermi wavenumber.

In the measurement scheme shown in the upper panel of Fig. 3(a), a backgate voltage $V_{bg}$ is applied to induce a band bending near the LSMO/STO interface, which in turn modifies the charge carrier density near the interface. Because the carrier (hole) density is fairly high in this film, the sheet resistance reveals only a small variation (≈0.02%) as $V_{bg}$ is swept from −40 to +40 V, Fig. 3(b). Nevertheless, the variation of $R_s$ with $V_{bg}$ reveals a noticeable non-monotonic feature. We can explain the non-monotonic behavior in terms of the presence of two conduction subbands, as depicted in Fig. 1(a). Because our film is clean, the heavy electrons in the spin-down subband are delocalized. The two conduction channels (light holes and heavy electrons) respond in an opposite way to the applied $V_{bg}$, and thus resulting in a non-monotonic characteristic of $R_s$ versus $V_{bg}$. This interpretation is supported by the Hall effect studies of LSMO, where a two-band model is needed to explain the data [19]. Moreover, this interpretation is supported by the angle-resolved photoemission studies [8]. In the latter experiment, a Fermi surface (FS) of an electron pocket centered around the Γ point was observed in metallic LSMO films, whereas the FS gradually diminished in those highly resistive samples lying close to the metal-insulator transition boundary [8]. The presence of delocalized spin-down electron states plays a crucial role for facilitating the WAL effect, as emphasized above [18].

Being a quantum-interference effect, the WAL MR is only important in the low magnetic field regime of $H \ll |H_c|$. To minimize any possible hysteretic effect on the measured WAL MR, we have symmetrized our data by taking the even part of $R_s(H)$, i.e., $R_s(H)_{\text{even}} = [R_s(H) + R_s(-H)]/2$. [For simplicity, in the following discussion we shall use the same notation $R_s(H)$ to denote the symmetrized data.] The positive parabolic background MR due to the Lorentz force (in high magnetic fields) has also been subtracted. Figure 3(c) shows the symmetrized, normalized MR measured at 0.36 K and at several $V_{bg}$ values. Note that the zero-field MR dip is sensitive to $V_{bg}$, reflecting a significant interfacial SOC effect due to a broken inversion symmetry at the LSMO/STO interface. (The hysteretic part of MR is also affected by $V_{bg}$. This issue requires further investigations.) The MR dip is the largest at $V_{bg} = 0$ V, with its magnitude decreasing with increasing $|V_{bg}|$. A gate tunable MR dip immediately implies that the SOC is of the Rashba type [20,21]. The Rashba SOC can nurture the D'yakonov-Perel' (DP) spin-relaxation processes [22,23], where spin relaxation arises from the spin precession between two consecutive elastic scattering events. The spin-relaxation, i.e., spin-orbit scattering, rate $\tau_{so}^{-1}$ is predicted to vary linearly with $\tau_e$. Therefore, the cleaner the system is made, the higher the $\tau_{so}^{-1}$ scattering rate will be [24,25]. Iordanskii–Lyanda-Geller–Pikus (ILP) have theoretically calculated the WAL MR by explicitly taking the DP spin-relaxation mechanism into account [10,26]. Their prediction for a quasi-two-dimensional (quasi-2D) system in the presence of a perpendicular magnetic field can be expressed in a compact form in terms of $\tau_\varphi$ and $\tau_{so}$ (Ref. [27]):

$$\frac{R_s(H) - R_s(0)}{R_s^2(0)} = -\frac{e^2}{2\pi^2 \hbar} \left\{ \Psi\left(\frac{1}{2} + \frac{H_\varphi + H_{so}}{H}\right) - \ln\frac{H_\varphi + H_{so}}{H} + \frac{1}{2}\Psi\left(\frac{1}{2} + \frac{H_\varphi + 2H_{so}}{H}\right) \right. \\ \left. - \frac{1}{2}\ln\frac{H_\varphi + 2H_{so}}{H} - \frac{1}{2}\Psi\left(\frac{1}{2} + \frac{H_\varphi}{H}\right) + \frac{1}{2}\ln\frac{H_\varphi}{H} \right\} \quad (1)$$

with

$$H_i \equiv \frac{\hbar}{4eD\tau_i}; \quad L_i = \sqrt{D\tau_i}, \quad i = \varphi, \text{so.}$$

where $\Psi(x)$ is the digamma function, and $H_\varphi$ is a characteristic scattering field. One can explicitly write $H_\varphi = \hbar/(4eD\tau_\varphi) = \hbar/(4eD\tau_{in}) + 2\hbar/(4eD\tau_s)$, where $\tau_{in}^{-1}$ is the total inelastic scattering rate, and $\tau_s^{-1}$ is the spin-spin (spin-flip) scattering rate due to magnetic impurities [18]. In practice, $\tau_s^{-1}$ is essentially temperature independent and can be extracted from the measured $\tau_\varphi^{-1}(T \to 0 \text{ K})$. We reiterate that in performing least-squares fits to Eq. (1), we have included only the MR data measured at $H < 0.5|H_c|$. Therefore, any possible hysteresis effect can largely be ignored. Also, in this low magnetic field regime, one does not need to consider the MR due to the many-body electron-electron interaction (EEI) effect [28].

The red curves in Fig. 3(c) are fitted curves using Eq. (1). Charge carrier dephasing length $L_\varphi$ and spin-orbit scattering length $L_{so}$ obtained from the fits are plotted as a function of $V_{bg}$ in Fig. 3(d). We find that $L_\varphi(V_{bg})$ takes a maximum value of $\approx 347$ nm at $V_{bg} = 0$, corresponding to an charge carrier dephasing time of $\tau_\varphi = 65$ ps and a characteristic scattering field of $H_\varphi = 1.4$ mT. Therefore, the WAL MR manifests at $H \ll |H_c|$. The obtained result $L_\varphi \gg t$ justifies the application of Eq. (1), where $t$ is the film thickness. We find that $L_\varphi$ ($\tau_\varphi$) decreases with increasing $|V_{bg}|$. This kind of $V_{bg}$ dependence is similar to what has previously been observed in topological insulators Bi$_2$Te$_3$ [29,30] and Bi$_2$Se$_3$ [29,30]. The applied gate voltage creates (enhances) an internal electrical field (which already exists at $V_{bg} = 0$ V) between the LSMO film and the backgate electrode, inducing a thin depletion (accumulation) layer by positive (negative) $V_{bg}$ for the spin-up conduction holes. Simultaneously, the applied gate voltage induces a thin accumulation (depletion) layer by positive (negative) $V_{bg}$ for the spin-down conduction electrons. Heuristically, we may envision the sample as constituted of a LSMO thick film (a bulk) and a thin interfacial LSMO/STO regime. Both are metallic and conduct in parallel. An applied $V_{bg}$ will have little effect on the electrical properties of the former, but can affect the latter regime markedly. Thus, the total $R_s$ of the sample, which is dominated by the bulk LSMO, shall depend only weakly on $V_{bg}$, as is seen Fig. 3(b). On the other hand, since $L_\varphi \gg t$ and the charge carriers traverse through the interfacial regime hundreds or thousands of times over $\tau_\varphi$, the quantum-interference quantities shall thus be modified by $V_{bg}$. Especially, if the dephasing processes are notably stronger in the interfacial regime than in the bulk LSMO, $L_\varphi$ and $L_{so}$ will reveal appreciable variations with $V_{bg}$. This is the case illuminated in Fig. 3(d). A stronger dephasing rate in the interfacial regime arises from the fact that this regime, being in proximity to a "dead" layer, is much less conductive compared to the bulk LSMO.

In weakly disordered metals ($k_F l_e > 1$), the magnitude and temperature dependence of $L_\varphi$ is determined by the responsible inelastic electron scattering processes. According to the current understanding, the carrier dephasing in 2D is governed by the Nyquist electron-electron (*e-e*) scattering at low temperatures, while the electron-phonon scattering can become important at somewhat higher temperatures [31,32]. In the EEI theory, the Coulomb interaction in a low-conductivity sample will be enhanced due to a suppression of the screening effect through the reduced density of states (or carrier concentration *n*) [28]. The *e-e* scattering rate is then increased, leading to a decreased $L_\varphi$ at low temperatures. In our case, upon the application of $V_{bg}$, a measurable increase in $R_s$ implies that the conductivity of the interfacial layer is drastically decreased. The phase-coherent charge carriers that traverse through this regime will then encounter a reduced *n* and *D* and undergo enhanced *e-e* (hole-hole) scattering. This picture explains why we have observed the $L_\varphi$ value to decrease with increasing $|V_{bg}|$.

Figure 3(d) shows that $L_{so}$ decreases monotonically from 72 to 59 nm as $V_{bg}$ increases from −20 to +40 V. The value of $L_{so}(V_{bg} = 0 \text{ V}) = 68.5$ nm corresponds to a spin-orbit scattering (spin-relaxation) rate of $\tau_{so}^{-1} \approx 3.9 \times 10^{11}$ s$^{-1}$. In the DP mechanism, $\tau_{so}^{-1} = (\Delta_{so}^2 \tau_e)/\hbar^2$, where $\Delta_{so}$ is the energy band splitting due to SOC [22]. We obtain $\Delta_{so}(V_{bg} = 0 \text{ V}) \approx 7.7$ meV from the above $\tau_{so}^{-1}$ value. The lower panel of Fig. 3(d) shows a monotonic increase of $\Delta_{so}$ with $V_{bg}$. From the relation $\Delta_{so} = 2k_F\alpha$ [33], we obtain the Rashba SOC coefficient $\alpha \approx 5.4 \times 10^{-13}$ eV m. For comparison, our $\Delta_{so}(V_{bg} = 0 \text{ V})$ value is on the same order of magnitude as that ($\approx 3.3$ meV) found in the 2DEG at the LaAlO$_3$/SrTiO$_3$ (LAO/STO) interface [24] and that ($\approx 5.4$ meV) in the InGaAs/InAlAs heterostructure [20]. Moreover, our value of $\Delta_{so}(V_{bg} = +40 \text{ V}) \approx 8.9$ meV is comparable to that [$\Delta_{so}(V_{bg} = +50 \text{ V}) \approx 7$ meV] found in the LAO/STO interface [24]. Due to possible uncertainties in the evaluations of $\tau_e$ and $D$, we estimate our extracted $\Delta_{so}$ value to be accurate to within a factor of ~2.

Our observation of backgate tunable $\Delta_{so}$ strongly suggests the presence of an internal electric field near the LSMO/STO interface. The electric field stems from a discontinuity in the layer-by-layer ionic structure of these two complex oxide materials, as has recently been theoretically predicted [34-36] and experimentally confirmed [37]. In this situation, the bandstructure in the interfacial regime will be different from that [Fig. 1(a)] in bulk LSMO. The interfacial bandstructure is to be modified by double exchange interaction [5,38] together with Rashba splitting, i.e., the two spin subbands split in both energy and momentum directions (see, for example, a schematic in Fig. 9 of [12]). Consequently, the $\Delta_{so}$ value should vary with $V_{bg}$ which adjusts the internal electric field near the interface.

Figure 4(a) shows the symmetrized, normalized MR at $V_{bg} = 0$ V and at several $T$ values between 0.36 and 10 K, as indicated. (The positive parabolic background MR has been subtracted, as mentioned.) The solid curves represent fitted curves using Eq. (1) and taking a constant $L_{so} = 68.5$ nm together with $L_\varphi$ as an adjusting parameter. The extracted $L_\varphi$ as a function of temperature is plotted in Fig. 4(b). The dashed straight line indicates a $T^{-1/2}$ temperature dependence and is a guide to the eye. At 10 K, the MR dip gradually evolves into a small MR peak (not shown), corresponding to a crossover from the strong SOC regime $L_\varphi > L_{so}$ to the weak SOC regime $L_\varphi < L_{so}$ at ~8 K. In short, the WAL effect is observed in our film which is clean and possesses a relatively large $|H_c| \gg H_\varphi$. A clean film with finite $\Delta_{so}$ ($\tau_{so}^{-1}$) renders a long dephasing length at low temperatures, and hence the criterion $L_\varphi > L_{so}$ is achieved.

Figure 4(c) shows the variation of $\tau_\varphi^{-1}$ with temperature. The black solid curve is a least-squares fit and will be discussed below. The dashed straight line is the theoretical prediction of the 2D Nyquist e-e scattering rate $\tau_{ee}^{-1}(T) = (A_{ee})^{th} T$, where $(A_{ee})^{th} = (e^2 k_B R_s / 2\pi \hbar^2) \times \ln(\pi \hbar / e^2 R_s)$, and $k_B$ is the Boltzmann constant [32]. From our $R_s$ value, we estimate the scattering strength to be $(A_{ee})^{th} = 8.5 \times 10^8$ s$^{-1}$ K$^{-1}$ for this sample. This theoretical value is more than one order of magnitude lower than the experimental value. However, the $R_s$ value in the interfacial regime should be much higher than the measured value. Therefore, the above calculation of $(A_{ee})^{th}$ is an underestimate (see below). Up to our highest measurement temperature of 10 K, no signature of electron-phonon scattering is observed, which should cause a $\tau_\varphi^{-1} \propto T^p$ dependence with the exponent $p \geq 2$ [31].

For comparison with the MR in the clean LSMO film, we have measured the MR in a 5-nm think film. The 5-nm thick film has a relatively high resistivity compared with that in the 22-nm thick film, see Fig. 1(b). Similar to previous results found in Refs. [11,17] and [11,17], our measurements reveal that the overall MR is negative in the wide magnetic field range $|H| \leq 1.2$ T, Fig. 5(a). Note that we do not observe any signature of the WAL effect even at $T < 2$ K. Instead, a zero-field peak in MR, i.e., the WL effect, is observed. Figure 5(b) shows the zero-field peaks for the 5-nm thick LSMO film at several $T$ values, as indicated. The magnitudes of the WL MR are small and nearly disappear at 5 K. The solid curves are the theoretical predictions of Eq. (1) with the fitting parameter $L_{so} \to \infty$, i.e., negligible SOC. The fitted $L_\varphi(T)$ values are 91, 85, 78, 64 and 35 nm at 0.36, 0.65, 1.0, 2.0 and 5.0 K, respectively. The inset of Fig. 5(b) shows the variation of the extracted $\tau_\varphi^{-1}$ with $T$ for this film. The solid curve is a least-squares fit to be discussed below.

To extract the spin-spin scattering rate due to magnetic impurities, we have fitted the measured $\tau_\varphi^{-1}(T)$ data in Figs. 4(c) and 5(b) to the form of $\tau_\varphi^{-1}(T) = 2\tau_s^{-1} + A_{ee} T$ [31]. The fitted values are $\tau_s^{-1} = 4.1 \times 10^9$ s$^{-1}$ and $A_{ee} = 3.1 \times 10^{10}$ s$^{-1}$ K$^{-1}$ for the 22-nm thick film, and $\tau_s^{-1} = 1.5 \times 10^{10}$ s$^{-1}$ and $A_{ee} = 3.8 \times 10^{10}$ s$^{-1}$ K$^{-1}$ for the 5-nm thick film. The 22-nm sample is much cleaner than the 5-nm sample in term of magnetic impurities. The magnetic impurities may (partly) originate from non-uniform magnetization near the LSMO/STO interface where the insulating "dead" layer can be antiferromagnetic [14,39]. With weak magnetic spin-spin scattering, the spin-relaxation processes are mainly dominated by the SOC processes in the 22-nm sample, with $\tau_{so}^{-1}/\tau_s^{-1} \simeq 100$. On the other hand, in the 5-nm thick sample, besides a larger $\tau_s^{-1}$, the spin-down subband states are probably (mostly) localized. Therefore, the SOC splitting $\Delta_{so}$ in this sample, if any exists, will be ineffective in causing spin relaxation, leading to $\tau_s^{-1} \gg \tau_{so}^{-1}$ and WL MR. Theoretically, Dugaev and coworkers have extended their original work [12] to show that WAL effect can happen in a ferromagnetic conductor under the condition $\tau_s^{-1} < \tau_{so}^{-1}$ [40]. They applied this new theory to explain the experimental observations by Neumaier *et al*. in (Ga,Mn)As nanostructures [41,42]. Our results for the 22-nm thick film also support this theoretical condition.

In the 5-nm thick film, we have $R_s = \rho_0/t = 560$ $\Omega$, giving rise to the *e-e* scattering strength $(A_{ee})^{th} \simeq 9 \times 10^9$ s$^{-1}$ K$^{-1}$. This is a factor of ~4 lower than the experimental value. Considering that there very likely exists a thin "dead" layer which will reduce the effective conduction thickness of the film, the discrepancy between experimental and theoretical values would be smaller. This level of consistency is satisfactory and supports that the inelastic dephasing processes in LSMO/STO is dominated by the 2D Nyquist *e-e* scattering. The fact that the fitted $A_{ee}$ value for the 22-nm film is similar to that for the 5-nm film suggests that the dephasing processes are much stronger in the interfacial LSMO/STO regime than in bulk LSMO.

## IV. CONCLUSION

We have studied a 22-nm thick epitaxial LSMO film with a low resistivity. We have observed the WAL MR, suggesting a marked SOC effect in this clean sample. The observations of non-monotonic gate-voltage dependent sheet resistance suggests an important contribution from the minority spin-down electrons to total electrical transport. The gate-voltage tunable SOC strength indicates the existence of an internal electric field near the LSMO/STO interface. Our extracted SOC splitting energy $\Delta_{so}$ is comparable to that found in the 2DEG at LAO/STO interface. The gate tunability of SOC may have potential applications in spintronic devices.

## Acknowledgments

J.J.L. was supported by Taiwan Ministry of Science and Technology through Grant number MOST-103-2112-M-009-017-MY3 and the Ministry of Education ATU Plan. M.Y. was supported by Grant-in-Aid for Young Scientists A (15H05517) from the Japan Society for the Promotion of Science (JSPS). H.O. was supported by Grant-in-Aid for Scientific Research on Innovative Areas (25106007) from JSPS. A part of this work was supported by Dynamic Alliance for Open Innovation Bridging Human, Environment and Materials.

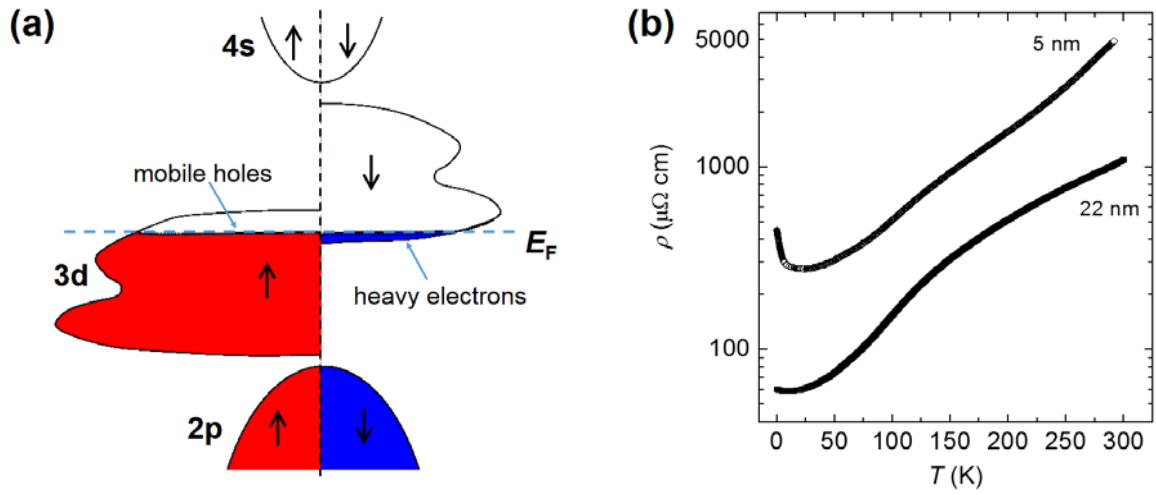

**FIG. 1.** (Color online) **(a)** Schematic energy bandstructure for a type III$_A$ half metal. $E_F$ denotes the Fermi energy. **(b)** Temperature dependence of resistivity $\rho(T)$ for a 22-nm and a 5-nm thick epitaxial LSMO films, as indicated.

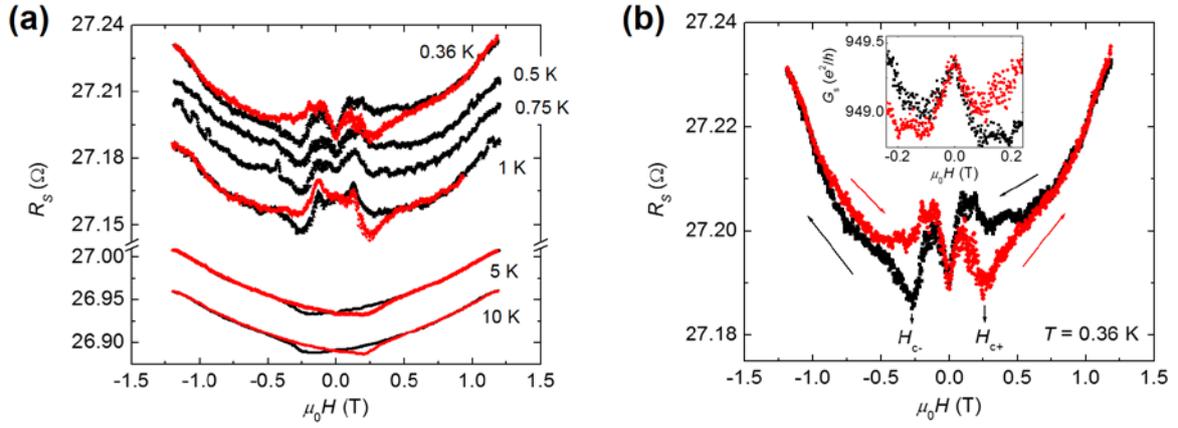

FIG. 2. (Color online) (a) MR at several $T$ values, as indicated. Black (red) curves correspond to the MR measured with magnetic field sweeping from +1.2 to −1.2 T (−1.2 to +1.2 T). (b) MR at $T = 0.36$ K. $H_{c+}$ and $H_{c-}$ denote the coercive fields. Inset: Sheet magneto-conductance $G_s(H)$ plotted in units of conductance quantum $e^2/h$.

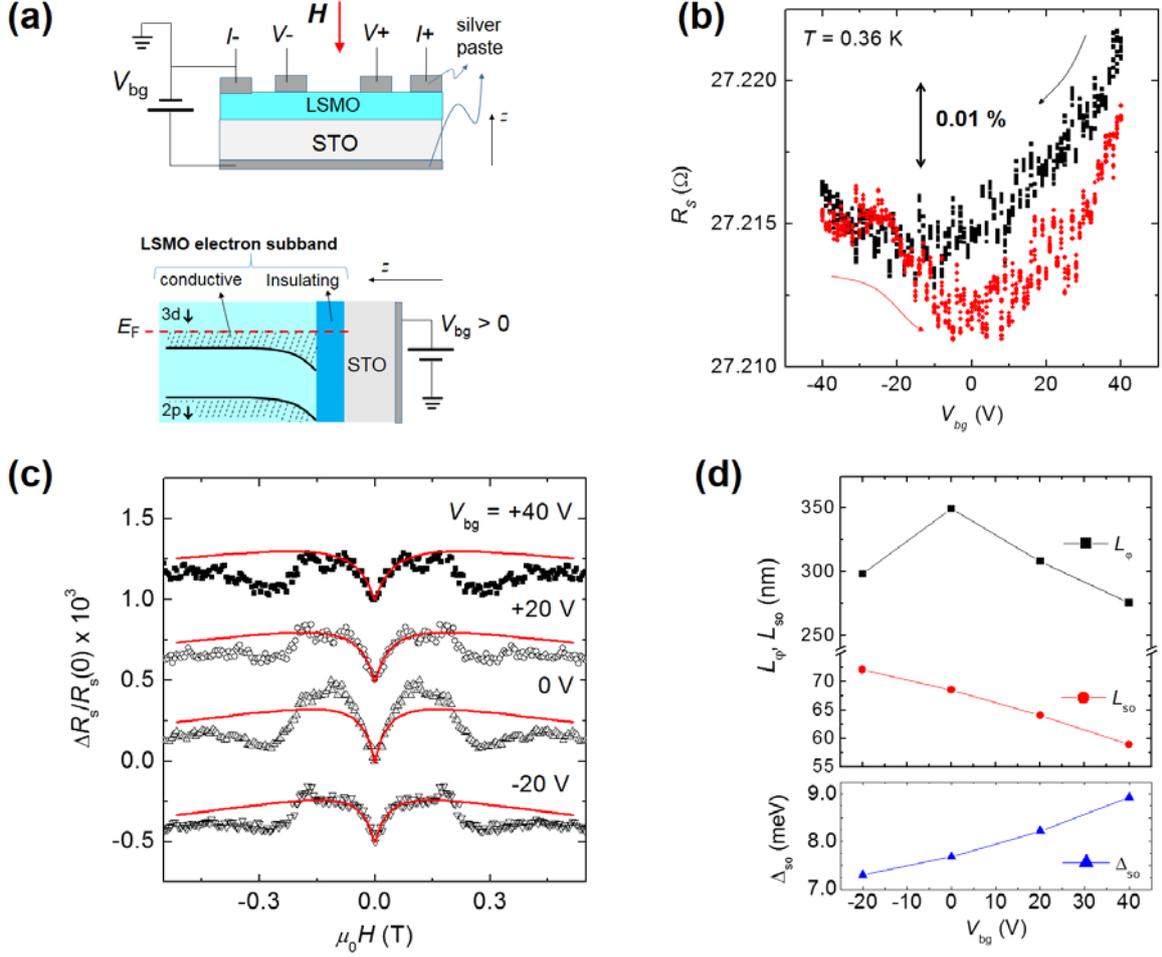

FIG. 3. (Color online) (a) Upper panel: a schematic of electrical measurement configuration. Lower panel: a schematic of electron band bending under a positive backgate voltage $V_{bg}$. (b) $R_s$ as a function of $V_{bg}$. The black (red) symbols are measured by sweeping $V_{bg}$ from +40 to −40 V (−40 to +40 V). (c) Symmetrized and normalized MR measured at 0.36 K and at several $V_{bg}$ values, as indicated. Data are offset for clarity. The solid curves are least-squares fits to Eq. (1). (d) Extracted dephasing length $L_\varphi$ and spin-orbit scattering length $L_{so}$ as a function of $V_{bg}$. The bottom panel shows the variation of SOC spin splitting energy $\Delta_{so}$ with $V_{bg}$.

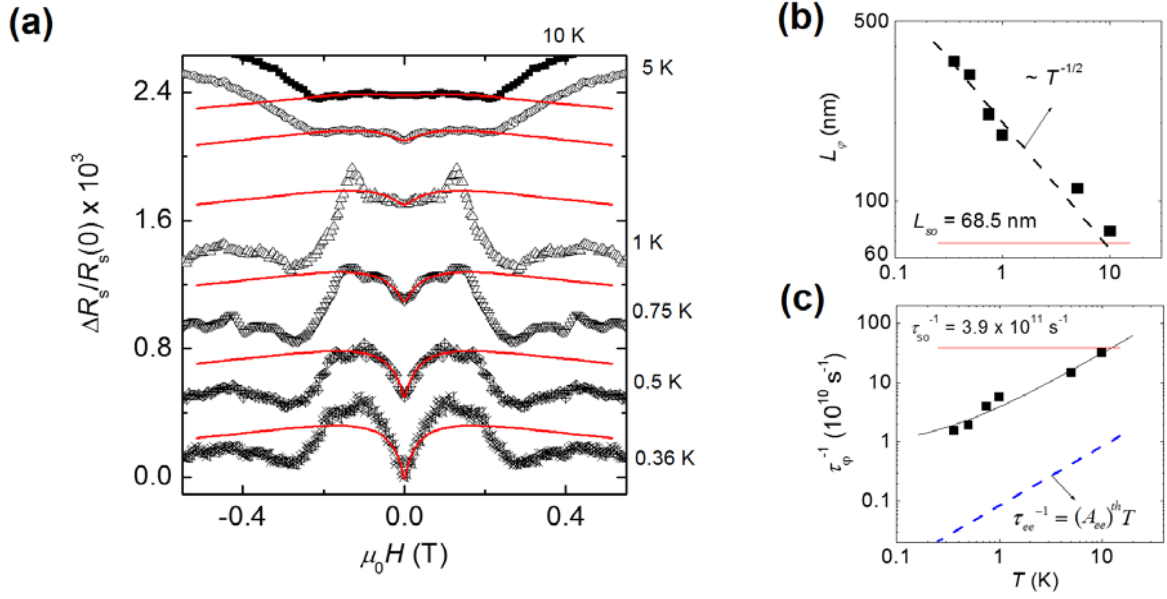

**FIG. 4.** (Color online) **(a)** Symmetrized and normalized MR measured at $V_{bg} = 0$ V and at several $T$ values, as indicated. Data are offset for clarity. The solid curves are least-squares fits to Eq. (1). **(b)** Extracted dephasing length $L_\varphi$ as a function of temperature. The dashed straight line indicates $T^{-1/2}$ temperature dependence and is a guide to the eye. The red horizontal line represents the spin-orbit scattering length $L_{so}$. **(c)** Dephasing rate $\tau_\varphi^{-1}$ as a function of temperature. The black solid curve is a least-squares fit to $\tau_\varphi^{-1} = 2\tau_s^{-1} + A_{ee}T$ (see text). The dashed straight line is the theoretical prediction of 2D *e-e* scattering rate $\tau_{ee}^{-1}(T)$, using the measured $R_s$ value. The red horizontal line represents the spin-orbit scattering rate $\tau_{so}^{-1}$.

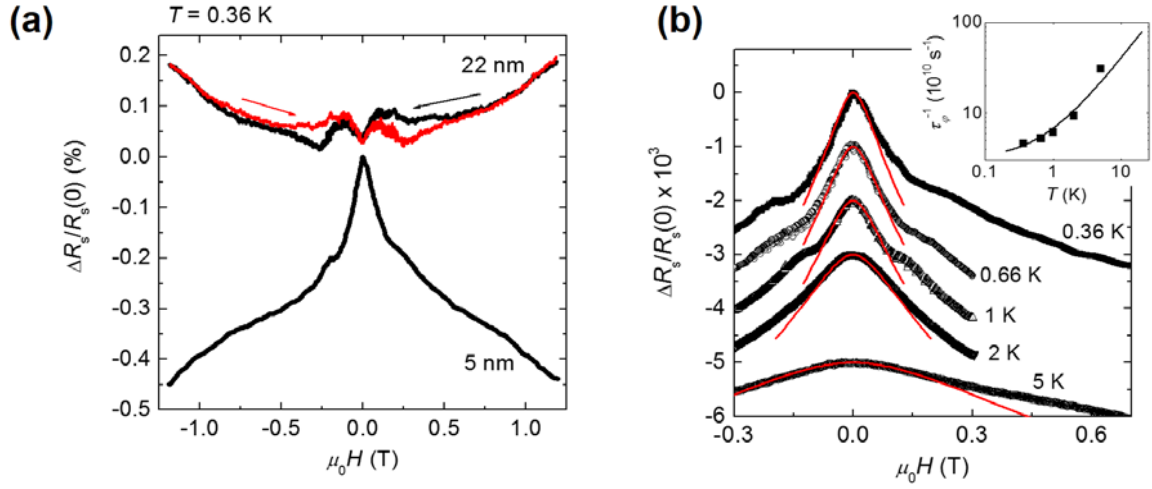

**FIG. 5.** (Color online) **(a)** MR at $T = 0.36$ K for a 5-nm and a 22-nm thick epitaxial LSMO films, as indicated. The MR for the 5-nm thick film was measured with magnetic field sweeping from +1.2 to −1.2 T. The data for the 22-nm thick film are taken from Fig. 2(a). **(b)** Low-field MR for the 5-nm thick LSMO film at several $T$ values, as indicated. The solid curves are least-squares fits to Eq. (1) with $\tau_{so}^{-1} = 0$. The inset show the variation of $\tau_\varphi^{-1}$ with temperature. The solid curve is a least-squares fit to $\tau_\varphi^{-1} = 2\tau_s^{-1} + A_{ee}T$ (see text).